\documentclass[aps,print,superscriptaddress,twocolumn]{revtex4-1}
\usepackage{amsfonts}
\usepackage{amsmath}
\usepackage{amssymb}
\usepackage{graphicx}
\usepackage{color}
\usepackage{float}
\makeatletter

\newcommand{\Rmnum}[1]{\expandafter\@slowromancap\romannumeral #1@}
\makeatother
\usepackage{dcolumn}
\usepackage{bm}
\usepackage[dvipsnames]{xcolor}
\usepackage{epsfig}
\usepackage{color}
\usepackage{graphics, graphicx}
\usepackage[toc,page]{appendix}
\begin{document}

\title{Fermi polaron in a dissipative bath with spin-orbit coupling}

\author{Jing Zhou}\email{zhjing@mail.ustc.edu.cn}
\affiliation{Department of Science, Chongqing University of Posts and Telecommunications,
Chongqing 400065, China}
\author{Wei Zhang}\email{wzhangl@ruc.edu.cn}
\affiliation{Department of Physics, Renmin University of China, Beijing 100872, China}
\affiliation{Beijing Key Laboratory of Opto-electronic Functional Materials and Micro-nano Devices,
Renmin University of China, Beijing 100872, China}

\date{\today}

\begin{abstract}
We study the polaron problem of an impurity immersed in a dissipative spin-orbit coupled Fermi gas via a non-self-consistent $T$-matrix method. We first propose an experimental scheme to realize a spin-orbit coupled Fermi bath with dissipation, and show that such a system can be described by a non-Hermitian Hamiltonian that contains an imaginary spin-flip term and an imaginary constant shift term. We find that the non-Hermiticity will change the single-particle dispersion of the bath gas, and modify the properties of attractive and repulsive polarons such as energy, quasi-particle residue, effective mass, and decay rate. We also investigate the Thouless criteria corresponding to the instability of the polaron--molecule transition, which suggests a molecule state is more facilitated with stronger bath dissipation. Finally, we consider the case with finite impurity density and calculate the interaction between polarons. Our result extends the study of polaron physics to non-Hermitian systems and may be realized in future experiment.
\end{abstract}

\maketitle

\section{Introduction}
Recently, the non-Hermitian system has attracted widespread attention from theorists and experimentalists. The non-Hermitian usually originates from the driven or dissipative processes induced by a bath, such as gain or loss of particle and energy from the environment. In principle, a strict approach to capture the whole characters of a driven and dissipative open system is to adopt the Lindblad equation. However, if we only consider short-time evolution where quantum jumps can be neglected, an effective non-Hermitian Hamiltonian can be approximately used to describe such systems. Non-Hermitian classical physics has been applied in many fields, for example, light propagating and scattering in a complex medium \cite{beenakker,Makris}, the friction of mechanical system \cite{Huber}, integrating resistor in an electrical circuit \cite{kottos}, and biological physics \cite{Celardo1,Celardo2}. Thanks to the recent developments in quantum technologies, non-Hermitian quantum physics plays a key role in understanding a vast of novel phenomena in quantum open systems. As a highly controllable platform, ultracold quantum gases of atoms can be used to implement many non-Hermitian Hamiltonian with laser-induced one-body \cite{Luo,Gadway,Gerbier} and two-body \cite{Tomita1,Tomita2} dissipation. Based on the experiment improvements, extensive theoretical works have analyzed non-Hermitian band theory \cite{Murakami}, topological phase transition \cite{hou}, novel magnetism \cite{song}, new linear response theory \cite{zhai} and non-Hermitian semimetal \cite{Sato}.

The concept of polaron is originally proposed by Landau and Pekar, and further elaborated by Fr\"{o}hlich and Feynman to describe the dressing effect of phonons on a Bloch electron. In contrast to an Anderson impurity, the impurity in polaron physics can move in the bath. Depending on the statistics obeyed by the bath, the polaron can be classified as Bose polaron and Fermi polaron, which have been both realized in experiment \cite{Chikkatur,Catani,Spethmann,Scelle, Schirotzek,Kohstall,Koschorreck} and analyzed theoretically via various methods. For Fermi polaron, Bishop \cite{Bishop} use perturbative expansion with interaction parameter $k_{F}a_{s}$ to investigate the repulsive polaron energy. A variational approach with particle-hole excitations~\cite{Chevy,Cui} is then proposed and extensively employed to treat the many-body effect in Fermi baths in different dimensionalities \cite{Zollner,parish1,parish2,Doggen,Kwasniok}, with spin-orbit coupling \cite{Yi}, near a narrow Feshbach resonance \cite{Massignan1,Qi}, and for an orbital Feshbach resonance \cite{chen1,chen2}. To further consider polaron decay, diagrammatic many-body method is implemented to give the polaron self-energy with ladder diagram approximation \cite{Combescot,Massignan2}. Fixed-node quantum Monte-Carlo (QMC) algorithm \cite{Prokofev1,Prokofev2,Vlietinck}, imaginary lattice quantum Monte-Carlo (ILMC) \cite{Bour}, functional renormalization group \cite{Schmidt}, and non-Gaussian variational method \cite{zhang1} have also been adopted to analyze this topic. However, impurity in a dissipative bath has not been studied so far to the best of our knowledge.

In this work, we consider an experimentally feasible non-Hermitian Fermi bath with spin-orbit coupling, and investigate the properties of a moving impurity immersed in this dissipative background via a non-self-consistent $T$-matrix method and effective Hamiltonian approximation. In particular, we obtain the polaron energy of both attractive and repulsive polaron, and characterize their properties by calculate the quasi-particle residue, the effective mass, and the two-body decay rate. To connect with experiment, we calculate the detectable spectrum function of the impurity atom to show the signal variation. We also discuss the Thouless criteria \cite{NSR} of pairing instability for attractive polaron branch, which suggests that a molecule state is more favorable in the strong dissipation regime. Finally, we extend our discussion to the case of finite impurity density. We calculate the variation of inter-polaron interaction energy with bath dissipation, which suggests the possibility of using dissipation as an extra controllable method in experiment.

The remainder of this paper is organized as follows: In Sec.~\ref{sec:system}, we describe an experimentally feasible system to realize a dissipative spin-orbit coupled Fermi system and derive an effective Hamiltonian, and discuss the single-particle dispersion spectrum. In Sec.~\ref{sec:formalism}, we use a non-self-consistent $T$-matrix theory to study the polaron problem in this system, and show the polaron energy and quasi-particle properties at zero temperature for a single impurity case. We also discuss the Thouless criteria of pairing instability to estimate the polaron--molecule transition. We then consider the finite impurity concentration and calculate the inter-polaron interaction versus bath dissipative strength in Sec.~\ref{sec:interaction}. Finally, we summarize the main results in Sec.~\ref{sec:conclusion}.

\section{Dissipative Fermi Bath With Spin-Orbit Coupling }
\label{sec:system}
\begin{figure}
\includegraphics[ width= 0.45\textwidth]{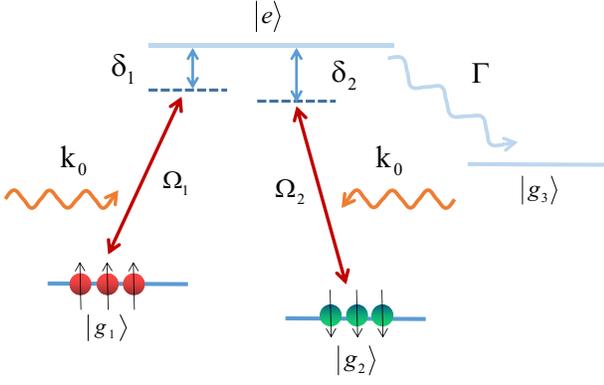}
\caption{(Color online) Four-energy level configuration of experimental realization of dissipative bath with synthetic spin-orbit coupling.}
\label{fig1}
\end{figure}
A dissipative spin-orbit coupled Fermi bath can be realized by a four-state scheme with three ground state energy levels $|g_{i}\rangle$ and an excited state $|e\rangle$ as shown in Fig.~\ref{fig1}. Two Raman lasers $\Omega_1$ and $\Omega_2$ shining along the $x$ direction with wave vector $k_0$ couple $|g_{1}\rangle$ and $|g_{2}\rangle$ to the excited state $|e\rangle$, with respective detuning $\delta_{1}$ and $\delta_{2}$. The excited state has a large decay rate $\Gamma$ to the third state $|g_{3}\rangle$. In order to describe the dissipative model, we introduce two Lindblad operators $S^{\pm}$ with the specific form shown in Appendix~\ref{append:lindblad}. Using the basis $\Phi^{T}=\{|g_{1}\rangle, |g_{2}\rangle, |g_{3}\rangle, |e\rangle\}$, the Raman coupling Hamiltonian is given by
\begin{equation}
H_{\rm Raman}=\left(
    \begin{array}{cccc}
      -\delta/{2} & 0 & 0 & \Omega_{1}^{\ast} \\
      0 & {\delta}/{2} & 0 & \Omega_{2}^{\ast} \\
      0 & 0 & 0 & 0 \\
      \Omega_{1} & \Omega_{2} & 0 & \Delta \\
    \end{array}
  \right).
\end{equation}
Here, we define $\delta=\delta_{1}-\delta_{2}$ and $\Delta=(\delta_{1}+\delta_{2})/2$ to simplify notation. The Lindblad equation for density matrix $\rho_s$ then takes the form
\begin{equation}
\frac{d\rho_{s}}{dt}=-i[H_{\rm Raman},\rho_{s}]+\Gamma \left[ S^{-}\rho_{s}S^{+}-\frac{1}{2}\left\{S^{+}S^{-},\rho_{s}\right\} \right],
\end{equation}
where $[\cdot, \cdot ]$ and $\{ \cdot , \cdot \}$ denote commutation and anti-commutation operations, respectively.

\begin{figure}
\includegraphics[width= 0.52\textwidth]{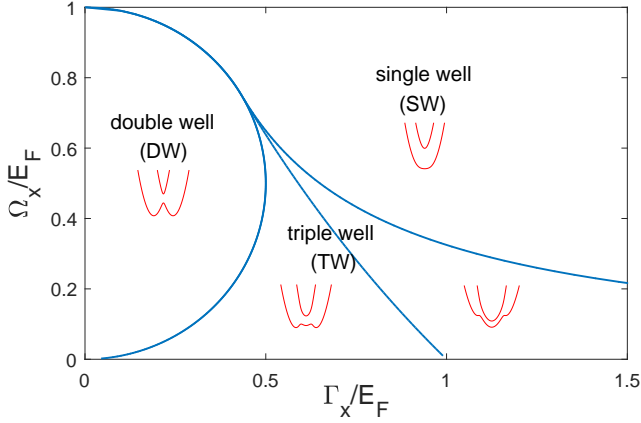}
\caption{(Color online) Phase diagram of single particle dispersion. Three types of dispersion structures can by identified on this diagram, which features respectively a single-well (SW), a double-well (DW) and a triple-well (TW) configuration. In this plot, we set $(k_{0}/k_F)^{2}=0.5$.}
\label{fig2}
\end{figure}

By getting the evolution of the elements of density matrix and adiabatically eliminating the excited state (details are shown in Appendix~\ref{append:lindblad}), we obtain the effective Hamiltonian of the spin-orbit coupled bath
\begin{equation}
H_{\rm bath}^{\rm eff}=\frac{(\mathbf{k}+k_{0}\textbf{e}_{x}\sigma_{z})^{2}}{2m}-\Omega_{x}\sigma_{x}
-i\Gamma_{x}(\sigma_{x}+\hat{I}).
\end{equation}
Here, we set the Raman coupling parameters $\Omega_{1}=\Omega_{2}=\Omega$ to simplify the model, and
use $\Omega_{x}=|\Omega|^{2} / \Delta$ and $\Gamma_{x}=\Gamma |\Omega|^{2} / \Delta^{2}$ to denote the spin flip strength and single particle dissipation, respectively. In the following discussion, we refer the ground levels $| g_1 \rangle$ and $| g_2 \rangle$ as pseudo-spin $| \uparrow \rangle$ and $| \downarrow \rangle$, respectively. We further assume the interaction between the ground states $|g_i\rangle$ is negligible, and the single-particle Hamiltonian can be diagonalized to reach the energy dispersion of the background
\begin{equation}
\varepsilon_{{\bf k}\pm}=\frac{\hbar^{2}(\textbf{k}^{2}+k_{0}^{2})}{2m}-i\Gamma_{x}\pm \sqrt{ \left(\frac{\hbar^{2}k_{x}k_{0}}{m}\right)^{2}+(\Omega_{x}-i\Gamma_{x})^{2}}.
\end{equation}
The phase diagram of the energy dispersion is shown in Fig.~\ref{fig2} with spin-orbit coupling parameter $(k_{0}/k_F)^{2}=0.5$ with $k_F$ the Fermi wavevector of the bath. In the following, we choose the natural unit $\hbar = m =1$, and set the Fermi energy $E_F$ as the energy unit. Three types of energy dispersion are observed by varying the parameters, including a single-well, double-well, and triple-well structures. The triple-well type can be further divided into two sub-categories by comparing the relative depths of the central and side energy minima. In the following discussion, we fix the Raman coupling $\Omega_x = E_F$ such that the single particle dispersion takes the single-well structure for all dissipation rate.

Then, we define the Matsubara Green's function of the bath $G_{\sigma\sigma'}^{(0)}=-\langle L| T_\tau C_{{\bf k}\sigma}(\tau)C_{{\bf k}\sigma'}^{\dag}(0)|R\rangle$, where $\langle L|$ and $|R\rangle$ are the left and right eigenvectors of the non-Hermitian Hamiltonian, $C$ and $C^\dagger$ are the fermionic operators, and ${T_\tau}$ is the time-ordering operator. This non-Hermitian Green's function can also be written in the following matrix form
\begin{equation}
G^{(0)}(\textbf{k},i\omega_{n})=
\left(
  \begin{array}{cc}
    G^{(0)}_{\uparrow\uparrow}(\textbf{k},i\omega_{n}) & G^{(0)}_{\uparrow\downarrow}(\textbf{k},i\omega_{n}) \\
    G^{(0)}_{\downarrow\uparrow}(\textbf{k},i\omega_{n}) & G^{(0)}_{\downarrow\downarrow}(\textbf{k},i\omega_{n}) \\
  \end{array}
\right),
\end{equation}
where the matrix elements are given by{\bf k}
\begin{align}
G^{(0)}_{\uparrow\uparrow}(\textbf{k},i\omega_{n})&=\frac{\psi_{{\bf k}+\uparrow}^{L}\psi_{{\bf k}+\uparrow}^{R}}{i\omega_{n}-\varepsilon_{{\bf k}+}}
+\frac{\psi_{{\bf k}-\uparrow}^{L}\psi_{{\bf k}-\uparrow}^{R}}{i\omega_{n}-\varepsilon_{{\bf k}-}}, \nonumber\\
G^{(0)}_{\uparrow\downarrow}(\textbf{k},i\omega_{n})&=\frac{\psi_{{\bf k}+\uparrow}^{L}\psi_{{\bf k}+\downarrow}^{R}}{i\omega_{n}-\varepsilon_{{\bf k}+}}
+\frac{\psi_{{\bf k}-\uparrow}^{L}\psi_{{\bf k}-\downarrow}^{R}}{i\omega_{n}-\varepsilon_{{\bf k}-}}, \nonumber\\
G^{(0)}_{\downarrow\uparrow}(\textbf{k},i\omega_{n})&=\frac{\psi_{{\bf k}+\downarrow}^{L}\psi_{{\bf k}+\uparrow}^{R}}{i\omega_{n}-\varepsilon_{{\bf k}+}}
+\frac{\psi_{k-\downarrow}^{L}\psi_{{\bf k}-\uparrow}^{R}}{i\omega_{n}-\varepsilon_{{\bf k}-}}, \nonumber\\
G^{(0)}_{\downarrow\downarrow}(\textbf{k},i\omega_{n})&=\frac{\psi_{{\bf k}+\downarrow}^{L}\psi_{{\bf k}+\downarrow}^{R}}{i\omega_{n}-\varepsilon_{{\bf k}+}}
+\frac{\psi_{{\bf k}-\downarrow}^{L}\psi_{{\bf k}-\downarrow}^{R}}{i\omega_{n}-\varepsilon_{{\bf k}-}}.
\end{align}
In the expressions above, $\psi_{k\nu\sigma}^{\lambda}$ is the transformation between the dressed-particle operator in the helix space and the original operator in the spin space, which is explained in detail in Appendix~\ref{append:Green}.

\section{properties of the polaron state}
\label{sec:formalism}

In this section, we consider a single impurity immersed in the dissipative Fermi bath with spin-orbit coupling as introduced in the previous section. The impurity is assumed to interact with one of the two ground levels (say, e.g., the $| \uparrow \rangle$ state) with a tunable strength by crossing a wide Feshbach resonance. The interaction Hamiltonian takes the $s$-wave contact potential form
\begin{equation}
H_{\rm int}=\frac{U}{V}\sum_{\textbf{k}\textbf{k}'\textbf{q}}C_{\textbf{q}/2+\textbf{k}\uparrow}^{\dag}
C_{\textbf{q}/2-\textbf{k}'\uparrow}b_{\textbf{q}/2-\textbf{k}}^{\dag}b_{\textbf{q}/2+\textbf{k}'},
\end{equation}
where $C_{\textbf{k}\uparrow}$ is the annihilation operator of the spin-up fermion, $b_{\textbf{k}}$ is the annihilation operator of the impurity, and $V$ is the quantization volume. We then use the many-body $T$-matrix theory to solve for the self-energy of the impurity. By keeping all the ladder-type diagrams, the self-energy at temperature $T$ is given as
\begin{equation}
\Sigma_{\rm tot}(\textbf{k},i\omega_{n})=k_{B}T\sum_{\textbf{q},i\Omega_{n}}G_{\uparrow\uparrow}^{(0)}(\textbf{q}-\textbf{k},i\Omega_{n}-i\omega_{n})\Gamma(\textbf{q},i\Omega_{n}),
\end{equation}
where the vertex function $\Gamma$ can be written through Bethe-Slapeter equation as
\begin{eqnarray}
&&\Gamma(\textbf{q},i\Omega_{n})^{-1}=\frac{1}{U}
\nonumber \\
&&\hspace{1cm}
+k_{B}T\sum_{\textbf{k},i\omega}
G^{(0)}(\textbf{k},i\omega)
G^{(0)}_{\uparrow\uparrow}(\textbf{q}-\textbf{k},i\Omega_{n}-i\omega).
\nonumber\\
\end{eqnarray}
Here, the free Green's function of the impurity takes the form $G^{(0)}({\bf k}, i\omega)={1}/{(i\omega-\epsilon_{\bf k}^I)}$ with impurity dispersion $\epsilon_{\bf k}^I$. Note that the vertex function has two parts which are contributed by the two helicity bands of the spin-orbit coupled Fermi background.

\begin{figure}
\centering
\includegraphics[width= 0.5\textwidth]{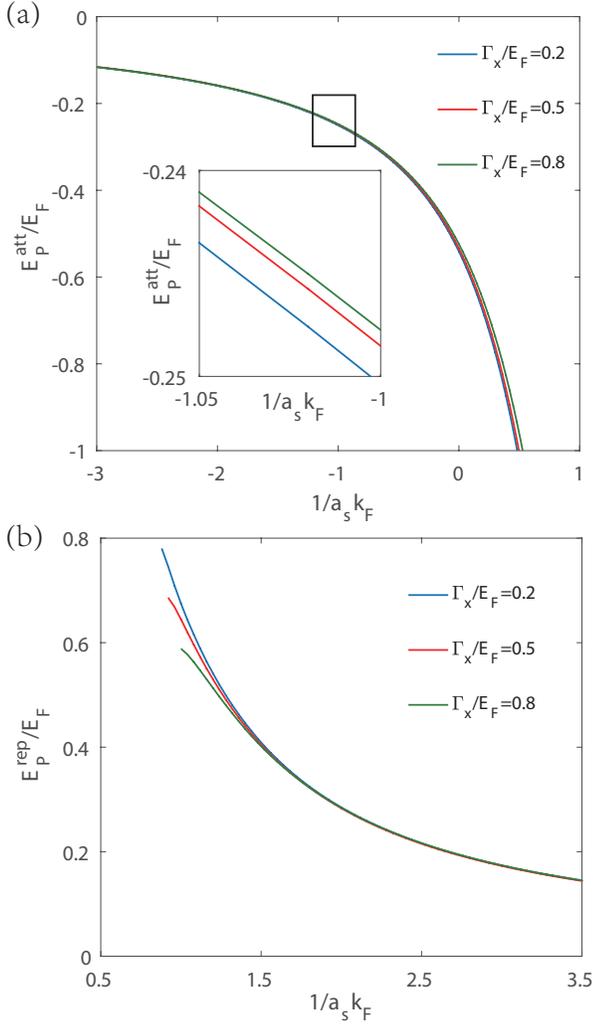}
\caption{(Color online) Polaron energy of the (a) attractive branch $E_{P}^{\rm att}$ and (b) repulsive branch $E_{P}^{\rm rep}$. Lines in all panels are plot with $\Gamma_{x}/E_{F}=0.2$ (blue), 0.5 (red), and 0.8 (green). We choose $(k_0/k_F)^2 = 0.5$ and $\Omega_x/E_F = 1$.}
\label{fig3}
\end{figure}

At zero temperature, after summing up the Matsubara frequency, the retarded self-energy is given by
\begin{align}
\Sigma_{\rm tot}^{R}(\textbf{k},i\Omega_{n})&=\frac{1}{V}\sum_{\textbf{q}} \bigg[ \Theta(-\varepsilon_{\textbf{q}+})\Phi_{+\uparrow}\Gamma^{R}
(\textbf{q}+\textbf{k},\varepsilon_{\textbf{q}+}+\omega^{+})\nonumber\\
&+\Theta(-\varepsilon_{\textbf{q}-})\Phi_{-\uparrow}\Gamma^{R}(\textbf{q}+\textbf{k},\varepsilon_{\textbf{q}-}+\omega^{+}) \bigg].
\end{align}
Here, $\Phi_{+\uparrow}=\psi_{\textbf{q}+\uparrow}^{L}\psi_{\textbf{q}+\uparrow}^{R}$, $\Phi_{-\uparrow}=\psi_{\textbf{q}-\uparrow}^{L}\psi_{\textbf{q}-\uparrow}^{R}$, and $\Theta(x)$ is the Heaviside step function as the zero-temperature limit of the Fermi-Dirac function. Owing to the non-Hermiticity of the spin-orbit coupled bath, there is an imaginary part in the dispersion $\varepsilon_{\textbf{k}\pm}$. However, we only consider the real part of the dispersion energy in the step function, because the imaginary part is connected with the life time of the dressed particle and only shows oscillation behaviors in the distribution function. We adopt a non-self-consistent ``$G_{0}G_{0}$" theory in the calculation which has been proved equivalent to the variational wavefunction approach \cite{Chevy}.

Once the self-energy is obtained, we can get the quasi-particle properties of the impurity from the retarded impurity Green's function
\begin{equation}
G_{I}^{R}(\textbf{k},\omega^{+})=\frac{1}{\omega-(\epsilon_{\bf k}-\mu_{I})-\Sigma_{\rm tot}^{R}(\textbf{k},\omega+i\eta^{+})+i\eta^{+}}
\end{equation}
where $\epsilon_{\bf k}$ is the impurity dispersion, and $\mu_I$ is the corresponding chemical potential.
In fact, within the quasi-particle approximation, the retarded impurity Green's function can also be expressed with quasi-particle ratio $Z$, effective mass $m^{\ast}_{\rm eff}$ and two-body decay rate $\gamma$. Considering the symmetry of the dispersion of the spin-orbit coupled bath, the dressed impurity would have two effective masses $m^{\ast}_x$ and $m^{\ast}_y=m^{\ast}_z=m^{\ast}_{||}$. Thus, in the low-energy and long-wavelength limit, the retarded impurity Green's function takes the form
\begin{eqnarray}
&&G_{I}^{R}(\textbf{k},\omega^{+})=
\nonumber \\
&&\frac{Z}{\omega-\hbar^{2}k_{||}^{2}/2m^{\ast}_{||}-\hbar^{2}k_{x}^{2}/2m^{\ast}_{x}+\mu_{I}-E_{P}+i\gamma/2}.
\end{eqnarray}
where, $k_{||}^{2}=k_{y}^{2}+k_{z}^{2}$. Compared with the two forms of retarded Green functions, the energy of the polaron state can be determined as
\begin{equation}
E_{P}={\rm Re} \Sigma_{\rm tot}^{R}(\textbf{k}=0,E_{P}-\mu_{I}),
\end{equation}
and the quasi-particle properties are characterized by
\begin{align}
Z&=\frac{1}{1-\frac{\partial {\rm Re} \Sigma_{\rm tot}^{R}}{\partial\omega}},\\
\frac{m}{m^{\ast}_{||}}&=\frac{1+\frac{\partial {\rm Re} \Sigma_{\rm tot}}{\partial\epsilon_{||}}}{1-\frac{\partial {\rm Re}\Sigma_{\rm tot}^{R}}{\partial\omega}}, \\
\frac{m}{m^{\ast}_{x}}&=\frac{1+\frac{\partial {\rm Re}\Sigma_{\rm tot}^{R}}{\partial\epsilon_{x}}}{1-\frac{\partial {\rm Re}\Sigma_{tot}^{R}}{\partial\omega}}, \\
\gamma&=-2Z {\rm Im}\Sigma_{\rm tot}^{R},
\end{align}
In the expressions above, $\epsilon_{||}$ and $\epsilon_{x}$ are impurity dispersions along $k_{||}$ and $k_x$, respectively.

\begin{figure}
\setlength{\belowcaptionskip}{0.0cm}
\includegraphics[width= 0.5\textwidth]{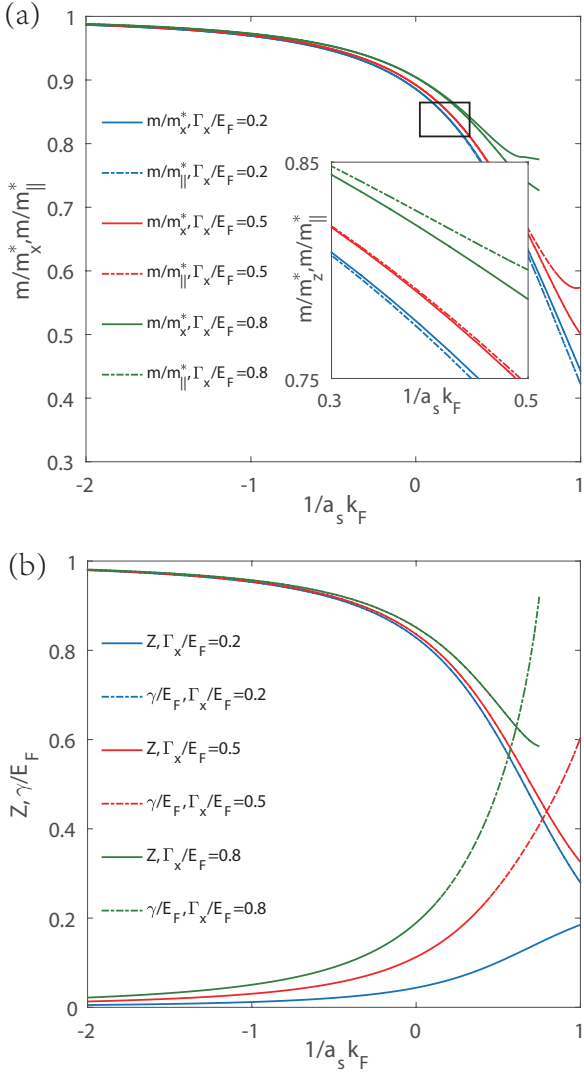}
\caption{(Color online) (a) Effective mass along the $k_{x}$ (solid) and $k_{II}$ (dashed-dotted) directions of the attractive polaron state. (b) Quasi-particle residue (solid) and two-body decay rate (dashed-dotted) of the attractive polaron state. Lines in all panels are plot with $\Gamma_{x}/E_{F}=0.2$ (blue), 0.5 (red), and 0.8 (green). Parameters are chosen to be same as in Fig.~\ref{fig3}.}
\label{fig4}
\end{figure}
In Fig.~\ref{fig3} and Fig.~\ref{fig4}, we show the energy, effective mass, quasi-particle residue, and two-body decay rate for polaron states with different bath dissipation strength. As depicted in subplots \ref{fig3}(a) and \ref{fig3}(b), the polaron energy increases with dissipation strength for the attractive polaron branch and decreases for the repulsive branch. The dependence is negligible in the deep Bardeen-Cooper-Schrieffer (BCS) limit for the attractive branch, and also in the Bose-Einstein condensate (BEC) limit for the repulsive branch, but becomes sizable around the unitary region. Owing to the presence of the one-dimensional spin-orbit coupling, the effective mass of the attractive polaron state acquires an anisotropy with different $m^{\ast}_{x}$ and $m^{\ast}_{||}$, as shown in Fig.~\ref{fig4}(a). The two effective masses both increase monotonically by crossing the Feshbach resonance from the BCS side to the BEC side, implying that the dressing effect is more significant on the impurity by the background with increasing interaction. On the other hand, the dissipation tends to make the impurity less inert and reduce the effective masses in all directions. An interesting finding is that for small dissipation, the polaron is easier to move along the $x$-direction with $m^{\ast}_{x} < m^{\ast}_{||}$. But the anisotropy inverses with increasing dissipation, showing a subtle competition between dissipation and anisotropic energy dispersion induced by spin-orbit coupling. In Fig.~\ref{fig4}(b), we plot the quasi-particle residue and two-body decay rate of the attractive polaron state. Notice that the impurity acquires larger quasi-particle residue with increasing dissipation, indicating that the impurity behaves more like an independent particle in that regime. This observation is qualitatively consistent with the trends shown in Figs.~\ref{fig3}(a) and \ref{fig4}(a). Finally, although the impurity does not have a direct decay channel, a larger bath dissipation will induce more severe decay of the attractive polaron state.

\begin{figure}
\includegraphics[width= 0.55\textwidth]{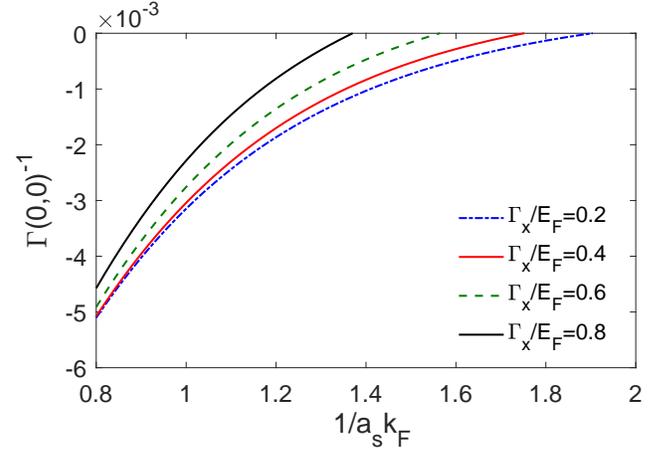}
\caption{(Color online) The Thouless criteria of the pairing instability for different dissipative strength. In this scheme, the polaron--molecule transition takes place at the point where $\Gamma(0,0)^{-1}$ reaches zero. Parameters are chosen to be same as in Fig.~\ref{fig3}.}
\label{fig5}
\end{figure}

Next, we investigate the polaron--molecule transition. Owing to the strong attractive interaction in the BEC regime, the impurity atom will tightly bounded with the bath fermion to form a dimer state. Such a transition point can be well described by the Thouless criteria of pairing instability $\Gamma^{-1}(\textbf{q}=0,i\Omega_{n}=0)=0$ for a non-dissipative Fermi system. However, in the present configuration of a dissipative bath, the Thouess criteria cannot be fully satisfied due to the presence of an imaginary part in the vertex function. Thus, we show in Fig.~\ref{fig5} only the real part of $\Gamma^{-1}(\textbf{q}=0,i\Omega_{n}=0)$, and neglect its imaginary part. We find that a larger bath dissipation tends to push the polaron--molecule transition point from the deep BEC limit towards the unitarity region, implying that the molecule state is more favorable with stronger dissipation.

\begin{figure}
\includegraphics[width= 0.55\textwidth]{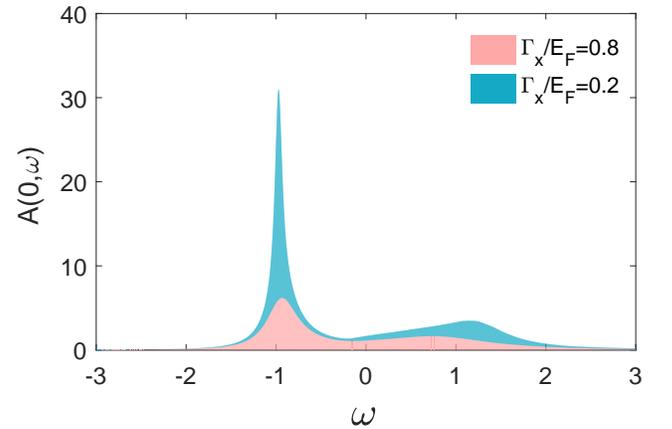}
\caption{(Color online) Spectrum function with different bath dissipation. Parameters are chosen to be same as in Fig.~\ref{fig3}.}
\label{fig6}
\end{figure}

To make a direct connection with experiments, we show in Fig.~\ref{fig6} the spectrum function which can be detected by spectroscopic measurement. Two peaks are observed and can be attributed respectively to the attractive and repulsive polaron states. By increasing the dissipative strength from $\Gamma_{x}/E_{F}=0.2$ to $\Gamma_{x}/E_{F}=0.8$ with scattering length $1/a_{s}k_{F}=0.5$ and SOC strength $(k_{0}/k_F)^{2}=0.5$, both peaks are shifted according to the results of Figs.~\ref{fig3}(a) and \ref{fig3}(b), and are significantly extended with smaller intensity owing to the stronger decay.

\section{Interaction between polarons}
\label{sec:interaction}

In this section, we consider the case of a finite impurity density to calculate the polaron-polaron interaction. Since the normal state of a highly imbalanced Fermi mixture can be understood as a Fermi liquid at zero temperature \cite{chevy_2010}, the ground state energy of this three components Fermi gases can be written in the form of the Landau-Pomeranchuk law as a function of the impurity concentration $x=n_{\rm imp}/n_{\rm bath}$
\begin{equation}
E=E_{\rm bath}+f(E_{b})x+g(m^{\ast}_{||}/m,m^{\ast}_{x}/m)x^{5/3}+Fx^{2},
\end{equation}
where $E_{\rm bath}$ is the kinetic energy of the non-interacting spin-orbit coupled bath. The second term comes from the binding energy of the impurity quasi particles in the background Fermi sea, the third term corresponds to the kinetic energy of the impurity atoms, and the last term is defined as the energy arising from the polaron-polaron interaction.
\begin{figure}
\includegraphics[width= 0.55\textwidth]{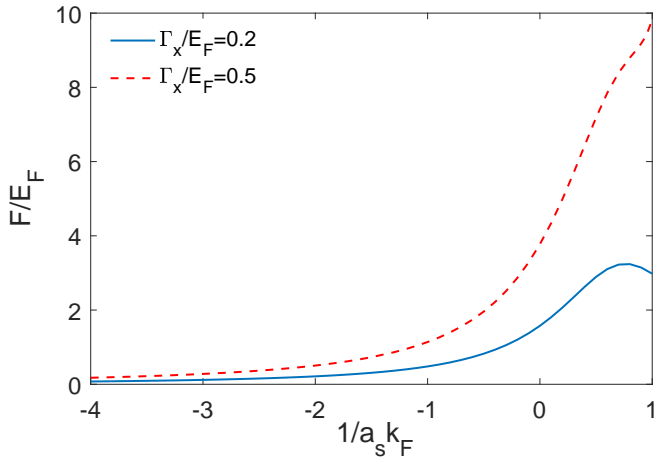}
\caption{(Color online) Polaron-polaron interaction parameter $F$ versus scattering length with different dissipation strengths. Parameters are chosen to be same as in Fig.~\ref{fig3}.}
\label{fig7}
\end{figure}
With the Gibbs-Duhem relation $\partial P/\partial \mu_{B}=n_{\rm bath}$ and $\partial P/\partial \mu_{I}=n_{\rm imp}$, we can obtain the grand-canonical equation of state
\begin{equation}
\label{eqn:EOS}
P=\int_{\min(E_{k-})}^{\mu_{B}}n_{\rm bath}(\mu)d\mu+\int_{0}^{\mu_{I}}n_{\rm imp}(\mu)d\mu.
\end{equation}
Here, $P$ is the pressure of the system, $\min (E_{k-})$ is the lowest energy band of the dissipative spin-orbit coupled bath, and $\mu_{B}$ and $\mu_{I}$ are the chemical potentials of the bath and the impurity, respectively. We need to emphasize that $\mu_{B}$ and $\mu_{I}$ are not the bare chemical potentials (Fermi energies) but include the contribution from the interaction. In order to get the total energy of the system, we convert the equation of state to the canonical ensemble, and get the relations of chemical potentials
\begin{align}
n_{\rm bath}(\mu_{B})&=n_{\rm bath} \left(1+x\frac{\partial E_{P}}{\partial\mu_{B}} \bigg|_{\mu_{B}=E_{BF}}\right),
\\
\mu_{I}&=E_{P}+E_{IF}.
\end{align}
Here, $E_{BF}={\hbar^{2}k_{BF}^{2}}/{2m}$ and $E_{IF}={\hbar^{2}k_{IF}^{2}}/{2m}$. From the first relation, we can have the renormalized bath chemical potential from the bare one (see Appendix~\ref{sec:append-EOS} for details). The second relation means that one has to cost a polaron energy $E_P$ plus an impurity Fermi energy $E_{IF}$ to add an impurity atom to the system with finite impurity concentration. Then, we arrive at the canonical ensemble energy function
\begin{equation}
E=-\sum_{i = B, I}P_{i}V_{i}+\sum_{i = B, I}\mu_{i}N_{i}.
\end{equation}
Rearranging the total energy $E$ in different powers of $x$ and fitting the coefficient of the $x^{2}$ term, we can get the polaron-polaron interacting parameter $F$ versus scattering length with different dissipative strength $\Gamma_{x}$. As illustrated in Fig.~\ref{fig7}, if the dissipative strength is small ($\Gamma_{x}/E_{F}=0.2$), the parameter $F$ increases first and then decreases by crossing the Feshbach resonance from the BCS to the BEC sides. However, if the dissipation is large enough, e.g., $\Gamma_{x}/E_{F}=0.5$, $F$ keeps increasing in the parameter region of scattering length considered here.

\section{conclusion and outlook}
\label{sec:conclusion}
In conclusion, we propose an experimentally feasible realization of a non-Hermitian spin-orbit coupled bath. Based on the effective Hamiltonian of the bath gases, we use non-self-consistent $T$-matrix theory to solve the polaron problem in this system. We obtain the variation of the polaron energy and quasi-particle parameters with different bath dissipation strength. Furthermore, we use Landau-Pomeranchuk energy to describe the system energy in the low impurity concentration \cite{chevy_2010}, and investigate the polaron-polaron interaction for different scattering lengths and bath dissipations.

We also notice that Wasak {\it et al.} have used Keldysh Green function method to solve the dissipative polaron and molecule states in the absence of spin-orbit coupling \cite{Piazza}. However, in our system, there is no direct dissipation for the impurity. The dissipation bath, therefore, provides a complex self-energy to the impurity atom, which is similar to the finite temperature polaron or the repulsive polaron states. So, we can omit the jump terms in the Lindblad equation, and use an effective non-Hermitian Hamiltonian and many-body $T$-matrix \cite{Combescot} to solve the polaron problem.

We acknowledge Wei Yi, Hui Hu, Jia Wang, Ran Qi, Wei Zheng, and JiaZhen Sun for helpful discussion.
This work is supported by the National Natural Science Foundation of China (Grants No. 11504038, No. 11522436, No. 11774425), the National Key R\&D Program of China (Grant No. 2018YFA0306501), the Beijing Natural Science Foundation (Z180013), the Joint fund of the Ministry of Education (6141A020333xx), and the Research Funds of Renmin University of China (Grants No. 16XNLQ03 and No. 18XNLQ15).

\appendix
\setcounter{equation}{5}
\renewcommand{\theequation}{\Alph{section}\arabic{equation}}
\setcounter{figure}{5}
\renewcommand{\thefigure}{A\arabic{figure}}

\begin{widetext}

\section{The Non-Hermitian Hamiltonian for a dissipative Fermi bath with spin-orbit coupling}
\label{append:lindblad}
The dynamics of the dissipative Fermi bath with spin-orbit coupling can be described by introducing two Lindblad operators in the internal state space $\{ |g_1 \rangle, |g_2 \rangle, |g_3\rangle, |e\rangle \}$
\begin{equation}
\begin{array}{cc}
  S^{-}=\left(
          \begin{array}{cccc}
           0 & 0 & 0 & 0 \\
           0 & 0 & 0 & 0 \\
           0 & 0 & 0 & 1 \\
           0 & 0 & 0 & 0 \\
         \end{array}
       \right),
  \quad   S^{+}=\left(
         \begin{array}{cccc}
           0 & 0 & 0 & 0 \\
           0 & 0 & 0 & 0 \\
           0 & 0 & 0 & 0 \\
           0 & 0 & 1 & 0 \\
         \end{array}
        \right).
\end{array}
\end{equation}
From the Lindblad master equation, we can obtain the time evolution equation of each element of the density operator
\begin{equation}
\rho^{\cdot}_{sij}=-i[H_{s},\rho_{s}]_{ij}+\Gamma[S^{-}\rho_{s}S^{+}-\frac{1}{2}\{S^{+}S^{-},\rho_{s}\}]_{ij}.
\end{equation}
where $[\cdot, \cdot]$ and $\{\cdot, \cdot\}$ denote commutation and anti-commutation operations, respectively. In the limiting case of large detuning, the population on the excited state and the evolution of the coherence elements between the two lower ground states $|g_{i=1,2}\rangle$ and the excited state $|e\rangle$ can be safely neglected, such that we can focus only on the Hilbert subspace consisted by the two ground states. After a straightforward calculation, we can get four time evolution equations for density matrix within this subspace
\begin{eqnarray}
\frac{d\rho_{11}}{dt}&=&-i\Omega^{\ast}_{1}\rho_{41}+i\Omega_{1}\rho_{14}, \label{eqn:rho11}\\
\frac{d\rho_{12}}{dt}&=&i\delta\rho_{12}-i\Omega^{\ast}_{1}\rho_{42}+i\Omega_{2}\rho_{14}, \label{eqn:rho12}\\
\frac{d\rho_{21}}{dt}&=&-i\delta\rho_{21}-i\Omega^{\ast}_{2}\rho_{41}+i\Omega_{1}\rho_{24},\label{eqn:rho21}\\
\frac{d\rho_{22}}{dt}&=&-i\Omega^{\ast}_{2}\rho_{42}+i\Omega_{2}\rho_{24}, \label{eqn:rho22}
\end{eqnarray}
and the four algebraic equations
\begin{eqnarray}
(\frac{\Gamma}{2}+i\Delta)\rho_{41}&=&-i\Omega_{1}\rho_{11}-i\Omega_{2}\rho_{21}, \label{eqn:rho41}\\
(\frac{\Gamma}{2}+i\Delta)\rho_{42}&=&-i\Omega_{1}\rho_{12}-i\Omega_{2}\rho_{22}, \label{eqn:rho42}\\
(\frac{\Gamma}{2}-i\Delta)\rho_{24}&=&i\Omega^{\ast}_{1}\rho_{21}+i\Omega^{\ast}_{2}\rho_{22}, \label{eqn:rho24}\\
(\frac{\Gamma}{2}-i\Delta)\rho_{14}&=&i\Omega^{\ast}_{1}\rho_{11}+i\Omega^{\ast}_{2}\rho_{12}. \label{eqn:rho14}
\end{eqnarray}
In the expressions above, we set the detuning $\delta_{1}=\delta_{2}=\Delta$ and $\delta_{1}-\delta_{2}=0$. We can solve $\rho_{41}$, $\rho_{14}$, $\rho_{42}$ and $\rho_{24}$ from Eqs.~(\ref{eqn:rho41})-(\ref{eqn:rho14}), and take the results back into the Lindblad equations ~(\ref{eqn:rho11})-(\ref{eqn:rho22}). Then, we can make the Markov approximation to omit the formal and back jump terms in the Lindblad equations, such that the dynamics of density matrix can be considered as that of an effective Hamiltonian $H_{\rm eff}=H_{s}-i\frac{1}{2}\sum_{i}L^{\dag}_{i}L_{i}$, which takes the following matrix form
\begin{eqnarray}
H_{11}^{\rm eff}&=&\frac{-i\Gamma|\Omega_{1}|^{2}/2}{\Delta^{2}+\Gamma^{2}/4}-\frac{\Delta|\Omega_{1}|^{2}}{\Delta^{2}+\Gamma^{2}/4},\\
H_{22}^{\rm eff}&=&\frac{-i\Gamma|\Omega_{2}|^{2}/2}{\Delta^{2}+\Gamma^{2}/4}-\frac{\Delta|\Omega_{2}|^{2}}{\Delta^{2}+\Gamma^{2}/4},\\
H_{12}^{\rm eff}&=&\frac{-i\Gamma\Omega^{\ast}_{1}\Omega_{2}/2}{\Delta^{2}+\Gamma^{2}/4}-\frac{\Delta\Omega^{\ast}_{1}\Omega_{2}}{\Delta^{2}+\Gamma^{2}/4},\\
H_{21}^{\rm eff}&=&\frac{-i\Gamma\Omega_{1}\Omega^{\ast}_{2}/2}{\Delta^{2}+\Gamma^{2}/4}-\frac{\Delta\Omega_{1}\Omega^{\ast}_{2}}{\Delta^{2}+\Gamma^{2}/4}.
\end{eqnarray}
In addition, since there is a photon momentum transfer when atoms are scattered in the Raman process, the Raman frequency becomes photon momentum dependent
\begin{equation}
\Omega_{1}=|\Omega_{1}|e^{-ik_{0}x}, \quad \Omega_{2}=|\Omega_{2}|e^{ik_{0}x}.
\end{equation}
Taking the new Raman couplings back into the effective Hamiltonian, and performing a unitary transformation
\begin{equation}
U=\left(
         \begin{array}{cc}
          e^{-ik_{0}x} & 0 \\
          0 & e^{ik_{0}x} \\
          \end{array}
         \right)
\end{equation}
to cancel the exponential terms, we can finally obtain the a non-Hermitian effective Hamiltonian for the bath
\begin{equation}
H_{\rm bath}^{\rm eff}=U(H_{\rm eff})U^{\dag} = \frac{(\textbf{k}-k_{0}\textbf{e}_{x}\sigma_{z})^{2}}{2m}-\frac{1}{\Delta}|\Omega|^{2}\sigma_{x}
-i\frac{\Gamma|\Omega|^{2}}{\Delta^{2}}(\sigma_{x}+\hat{I}).
\end{equation}
In the main text, we define the effective magnetic field along the $x$ direction $\Omega_{x}=|\Omega|^{2} /\Delta$ and the dissipative strength $\Gamma_{x}= \Gamma|\Omega|^{2} / \Delta^{2}$, which can be adjusted independently.


\section{The free Green's function of the non-Hermitian spin-orbit coupled bath}
\label{append:Green}
Owing to the non-Hermiticity of bath effective Hamiltonian, we need to use left vector and right vector method to get the eigenvectors. We introduce two transform matrices $X_{L}$ and $X_{R}$ from the spin space to the eigenstate space of helix
\begin{equation}
\begin{array}{cc}
  \left(
  \begin{array}{c}
    C_{{\bf k}\uparrow} \\
    C_{{\bf k}\downarrow} \\
  \end{array}
\right) \equiv
X_{L}   \left(
                 \begin{array}{c}
                   C_{{\bf k}+}^{L} \\
                   C_{{\bf k}-}^{L} \\
                 \end{array}
               \right)
=
\left(
          \begin{array}{cc}
            \psi_{{\bf k}+\uparrow}^{L} & \psi_{{\bf k}-\uparrow}^{L} \\
            \psi_{{\bf k}+\downarrow}^{L} & \psi_{{\bf k}-\downarrow}^{L} \\
          \end{array}
        \right)\left(
                 \begin{array}{c}
                   C_{{\bf k}+}^{L} \\
                   C_{{\bf k}-}^{L} \\
                 \end{array}
               \right), \quad
\left(
  \begin{array}{c}
    C_{{\bf k}\uparrow}^{\dag} \\
    C_{{\bf k}\downarrow}^{\dag} \\
  \end{array}
\right) \equiv
X_{R}   \left(
                 \begin{array}{c}
                   C_{{\bf k}+}^{R} \\
                   C_{{\bf k}-}^{R} \\
                 \end{array}
               \right)
=
\left(
          \begin{array}{cc}
            \psi_{{\bf k}+\uparrow}^{R} & \psi_{{\bf k}-\uparrow}^{R} \\
            \psi_{{\bf k}+\downarrow}^{R} & \psi_{{\bf k}-\downarrow}^{L} \\
          \end{array}
        \right)\left(
                 \begin{array}{c}
                   C_{{\bf k}+}^{R} \\
                   C_{{\bf k}-}^{R} \\
                 \end{array}
               \right),
\end{array}
\end{equation}
where $C_{{\bf k},\sigma}$ and $C_{{\bf k}, \alpha}$ are fermionic operators of spin $\sigma = \uparrow\downarrow$ and helix $\alpha = \pm$, respectively. The coefficients in the transformation matrices can be determined by the biorthogonal condition  $X_{R}X_{L}=\hat{I}$ and the normalization relations $\sum_{\mu}\overline{\psi}^{R}_{{\bf k}\mu\uparrow}\psi^{R}_{{\bf k}\mu\uparrow}=1$ and $\sum_{\mu}\overline{\psi}^{R}_{{\bf k}\mu\downarrow}\psi^{R}_{{\bf k}\mu\downarrow}=1$ of $X_R$.

With the coefficient matrices, we can derive the free Green's function of the spin-orbit coupled bath, which is defined as
\begin{equation}
G^{(0)}(\textbf{k},\tau)=\left(
                           \begin{array}{cc}
                             -\langle T_{\tau}C_{{\bf k}\uparrow}(\tau)C_{{\bf k}\uparrow}^{\dag}(0)\rangle & -\langle T_{\tau}C_{{\bf k}\uparrow}(\tau)C_{{\bf k}\downarrow}^{\dag}(0)\rangle \\
                             -\langle T_{\tau}C_{{\bf k}\downarrow}(\tau)C_{{\bf k}\uparrow}^{\dag}(0)\rangle & -\langle T_{\tau}C_{{\bf k}\downarrow}(\tau)C_{{\bf k}\downarrow}^{\dag}(0)\rangle \\
                           \end{array}
                         \right)
\end{equation}
with $T_\tau$ the time-ordering operator.
The element $G_{\sigma\sigma'}$ takes the form
\begin{eqnarray}
G_{\sigma\sigma'}^{(0)}({\bf k},\tau)&=&-\langle T_{\tau}C_{{\bf k}\sigma}(\tau)C_{{\bf k}\sigma'}^{\dag}(0)\rangle\nonumber\\
&=&-\Theta(\tau)\psi_{{\bf k}+\sigma}^{L}\psi_{{\bf k}+\sigma'}^{R}e^{-\varepsilon_{{\bf k}+}\tau}(1-n_{{\bf k}+})
-\Theta(\tau)\psi_{{\bf k}-\sigma}^{L}\psi_{{\bf k}-\sigma'}^{R}e^{-\varepsilon_{{\bf k}-}\tau}(1-n_{{\bf k}-}),
\end{eqnarray}
where $\Theta(\tau)$ is the Heaviside step function, $\varepsilon_{{\bf k} \alpha}$ is the dispersion of helix with branch index $\alpha$, and $n_{{\bf k}\alpha}$ is the corresponding Fermi distribution.
Transforming to the frequency space, we obtain the final expression
\begin{eqnarray}
G_{\sigma\sigma'}^{(0)}({\bf k},i\omega_{n})&=&\int_{0}^{\beta}G_{\sigma\sigma'}^{(0)}({\bf k},\tau)e^{i\omega_{n}\tau}d\tau\nonumber\\
&=&\frac{\psi_{{\bf k}+\sigma}^{L}\psi_{{\bf k}+\sigma'}^{R}}{i\omega_{n}-\varepsilon_{{\bf k}+}}+
\frac{\psi_{{\bf k}-\sigma}^{L}\psi_{{\bf k}-\sigma'}^{R}}{i\omega_{n}-\varepsilon_{{\bf k}-}}.
\end{eqnarray}

\section{The impurity self-energy}
\label{sec:append-selfenergy}

To calculate the self-energy of the impurity, we first emphasize that there is no direct dissipative process of the impurity. So we can use Matsabara Green's function to calculate the self-energy directly.
With the ladder-diagram approximation, the vertex function is given by
\begin{eqnarray}
\Gamma^{-1}(\textbf{q},i\Omega_{n})&=&\frac{1}{U}+k_{B}T\sum_{\textbf{k},i\omega}
G^{(0)}(\textbf{k},i\omega)
G^{(0)}_{\uparrow\uparrow}(\textbf{q}-\textbf{k},i\Omega_{n}-i\omega)\nonumber\\
&=&\frac{1}{U}-\frac{1}{V}\sum_{k}(\frac{\psi_{\textbf{q}-\textbf{k},+\uparrow}^{L}\psi_{\textbf{q}-\textbf{k},+\uparrow}^{R}\Theta(\varepsilon_{\textbf{q}-\textbf{k},+})}
{i\Omega_{n}-\varepsilon_{\textbf{q}-\textbf{k},+}-\epsilon^{I}_{k}}+\frac{\psi_{\textbf{q}-\textbf{k},-\uparrow}^{L}\psi_{\textbf{q}-\textbf{k},-\uparrow}^{R}\Theta(\varepsilon_{\textbf{q}-\textbf{k},-})}
{i\Omega_{n}-\varepsilon_{\textbf{q}-\textbf{k},-}-\epsilon^{I}_{k}})
\end{eqnarray}
where $G^{(0)}({\bf k}, i\omega)={1}/{(i\omega-\epsilon_{\bf k}^I)}$ is the free Green's function of the impurity with dispersion $\epsilon_{\bf k}^I$. Then, the self-energy is
\begin{eqnarray}
\Sigma_{\rm tot}(\textbf{k},i\omega_{n})&=&k_{B}T\sum_{\textbf{q},i\Omega_{n}}G^{(0)}_{\uparrow\uparrow}(\textbf{q}-\textbf{k},i\Omega_{n}-i\omega_{n})\Gamma(\textbf{q},i\Omega_{n})\nonumber\\
&=&k_{B}T\sum_{q,i\Omega_{n}} \left( \frac{\psi_{\textbf{q}-\textbf{k},+\uparrow}^{L}\psi_{\textbf{q}-\textbf{k},+\uparrow}^{R}}{i\Omega_{n}-i\omega_{n}-\varepsilon_{\textbf{q}-\textbf{k},+}}
+\frac{\psi_{\textbf{q}-\textbf{k},-\uparrow}^{L}\psi_{\textbf{q}-\textbf{k},-\uparrow}^{R}}{i\Omega_{n}-i\omega_{n}-\varepsilon_{\textbf{q}-\textbf{k},-}}   \right)\Gamma({\bf q},i\Omega_{n})
\end{eqnarray}
Here, we don't consider the poles in the vertex function, which correspond to the molecule state. We then sum up the Matsabara frequency $\Omega_{n}$, and get the retarded self-energy
\begin{eqnarray}
\Sigma_{\rm tot}^{R}(\textbf{k},\omega^{+})=\frac{1}{V}\sum_{{\bf q}} \left[\Theta(-\varepsilon_{{\bf q}+})\psi_{\textbf{q},+\uparrow}^{L}\psi_{\textbf{q},+\uparrow}^{R}
\Gamma^{R}({\bf q+k},\varepsilon_{{\bf q}+}+\omega^{+})+\Theta(-\varepsilon_{{\bf q}-})\psi_{\textbf{q},-\uparrow}^{L}\psi_{\textbf{q},-\uparrow}^{R}
\Gamma^{R}({\bf q+k},\varepsilon_{{\bf q}-}+\omega^{+})
\right].
\end{eqnarray}
The vertex functions for the two branches have the form
\begin{eqnarray}
\Gamma^{R-1}({\bf q+k},\varepsilon_{{\bf q}+}+\omega^{+})=\frac{1}{U}-\frac{1}{V}
\left[ \frac{\psi_{{\bf k}_{1}+\uparrow}^{L}\psi_{{\bf k}_{1}+\uparrow}^{R}\Theta(\varepsilon_{{\bf k}_{1}+})}{\varepsilon_{{\bf q}+}+\omega^{+}-\varepsilon_{{\bf k}+}-\epsilon_{{\bf q+k-k_{1}}}^{I}}
+\frac{\psi_{{\bf k_{1}}+-\uparrow}^{L}\psi_{{\bf k_{1}}-\uparrow}^{R}\Theta(\varepsilon_{{\bf k_{1}}-})}{\varepsilon_{{\bf q}+}+\omega^{+}-\varepsilon_{{\bf k}-}-\epsilon_{{\bf q+k-k_{1}}}^{I}} \right],\\
\Gamma^{R-1}({\bf q+k},\varepsilon_{{\bf q}-}+\omega^{+})=\frac{1}{U}-\frac{1}{V}
\left[\frac{\psi_{{\bf k_{1}}+\uparrow}^{L}\psi_{k_{1}+\uparrow}^{R}\Theta(\varepsilon_{k_{1}+})}{\varepsilon_{{\bf q}-}+\omega^{+}-\varepsilon_{{\bf k}+}-\epsilon_{{\bf q+k-k_{1}}}^{I}}
+\frac{\psi_{{\bf k_{1}}+-\uparrow}^{L}\psi_{{\bf k_{1}}-\uparrow}^{R}\Theta(\varepsilon_{{\bf k_{1}}-})}{\varepsilon_{{\bf q}-}+\omega^{+}-\varepsilon_{{\bf k}-}-\epsilon_{{\bf q+k-k_{1}}}^{I}} \right].
\end{eqnarray}

\section{The equation of state}
\label{sec:append-EOS}

The chemical potential of the impurity atoms with finite density can be written as
\begin{equation}
\mu_{I}=E_{P}+E_{IF},
\end{equation}
which means we need to spend a polaron state energy plus an impurity Fermi energy to add another impurity atom to the system with finite impurity concentration. So, we begin from calculating the Fermi energy of the minority atoms
\begin{eqnarray}
\frac{k_{IF}(\theta) \sin^{2}\theta}{2m_{||}^{\ast}}+\frac{k_{IF}(\theta) \cos^{2}\theta}{2m_{x}^{\ast}}=E_{IF},\\
\sum_{\bf k}1=\frac{1}{(2\pi)^{2}}\int_{0}^{1}ds\int_{0}^{k_{IF}(\theta)}k^{2}dk=n_{\rm imp},
\end{eqnarray}
where $k_{IF}(\theta)$ is the Fermi wavevector at the azimuthal angle $\theta$ about the $x$-axis, and $s=\cos\theta$. By carrying out the integration, we can get $E_{IF}={(6\pi^{2}n_{\rm imp})^{2/3}}/{A^{2/3}}$ with
\begin{equation}
A=\int_{0}^{1} {\left(\frac{1-s^{2}}{2m_{||}^{\ast}}+\frac{s^{2}}{2m_{s}^{\ast}}\right)^{-3/2}}ds.
\end{equation}
Meanwhile, the Gibbs-Duhem relation leads to the following identities
\begin{equation}
  \frac{\partial P}{\partial\mu_{B}}=n_{\rm bath},  \quad  \frac{\partial P}{\partial\mu_{I}}=n_{\rm imp},
\end{equation}
where $\mu_{B, I}$ are chemical potentials of the bath (B) and and impurities (I), respectively.
By integrating $\mu_{B}$ and $\mu_{I}$, we can get the expression for pressure
\begin{equation}
P=\int_{\min(E_{k-})}^{\mu_{B}}n_{\rm bath}(\mu)d\mu+\frac{A}{15\pi^{2}}(\mu_{I}-E_{P})^{5/2}.
\end{equation}
From $\frac{\partial P}{\partial \mu_{B}}=n_{\rm bath}$, we can get
\begin{equation}
n_{\rm bath}=n_{\rm bath}(\mu_{B})-n_{\rm imp} \left(\frac{\partial E_{P}}{\partial\mu_{B}}\right) \bigg|_{\mu_{
B}=E_{BF}}.
\end{equation}
With the definition of impurity concentration $x={n_{\rm imp}}/{n_{\rm bath}}$, we can rewrite the equation above as
\begin{equation}
n_{\rm bath}(\mu_{B})=n_{\rm bath} \left[ 1+x \left(\frac{\partial E_{P}}{\partial\mu_{B}}\right) \bigg|_{\mu_{B}=E_{BF}} \right].
\label{eqn:renormalize}
\end{equation}
From this equation, we can see $n_{\rm bath}(\mu_{B})=n_{\rm bath}$ if there is only one impurity, which reduces to the case of conventional polaron problem. However, if the density of impurity atoms is finite, the density of the majority bath gases needs be renormalized as shown in Eq.~(\ref{eqn:renormalize}).
Till now, we can get the total energy of the system as:
\begin{eqnarray}
E/V&=&-\sum_{i=B,I}P_{i}+\sum_{i=B,I}\mu_{i}n_{i}\nonumber\\
&=&-\int_{\min(E_{k-})}^{\mu_{B}}n_{\rm bath} \left[ 1+x \left(\frac{\partial E_{P}}{\partial\mu_{B}}\right) \bigg|_{\mu_{B}=E_{BF}} \right] d\mu
- \frac{A}{15\pi^{2}} \left(\frac{6\pi^{2}n_{\rm imp}}{A}\right)^{5/3}+\mu_{B}n_{\rm bath}(\mu_{B})\nonumber\\
&& \hspace{1cm}
+\left[ \left(\frac{\partial E_{P}}{\partial\mu_{B}}\right) \bigg|_{\mu_{B}=E_{BF}}\mu_{B}+\left(\frac{6\pi^{2}n_{\rm imp}}{A}\right)^{2/3} \right]n_{\rm imp}.
\end{eqnarray}
\end{widetext}


\end{document}